# An Approach for Computing Dynamic Slice of Concurrent Aspect-Oriented Programs

Abhishek Ray[1], Siba Mishra[1] and Durga Prasad Mohapatra[2]

[1] School of Computer Engineering, Kiit University, Bhubaneswar, Odisha, India
[2] Department of Computer Science & Engineering, National Institute of Technology, Rourkela, Odisha, India
ar_mmclub@yahoo.com, sibamishra@yahoo.co.in, durga@nitrkl.ac.in

*Abstract*

*We propose a dynamic slicing algorithm to compute the slice of concurrent aspect-oriented programs. We use a dependence based intermediate program representation called Concurrent Aspect-oriented System Dependence Graph (CASDG) to represent a concurrent aspect-oriented program. The CASDG of an aspect-oriented program consists of a system dependence graph (SDG) for the non-aspect code, a group of dependence graphs for aspect code and some additional dependence edges used to connect the system dependence graph for the non-aspect code to dependence graph for aspect code. The proposed dynamic slicing algorithm is an extended version of NMDS algorithm for concurrent object-oriented programs, which is based on marking and unmarking of the executed nodes in CASDG appropriately during run-time.*

## 1. Introduction

Program slicing [3, 4, 7, 14, 28, 29, 30, 31] is a practical disintegration methodology that omits program modules that are irrelevant to a particular computation process based on a criterion known as the slicing criterion [29, 30, 31]. The original program's semantics is projected through the computation of an executable program formed by the left over modules called a *slice*. Using this methodology, we can automatically determine the relevance of a module in a particular computation. The concept of *static program slicing* was introduced by Weiser [29, 30, 31]. A program $P$ can be sliced with respect to a slicing criterion. A *slicing criterion* [29, 30, 31] consists of a pair $< s, v >$, where $s$ is a program point and $v$ is a subset of program variables. The parts of a program that have a direct or indirect effect on the values computed at a slicing criterion are called the program slice with respect to criterion. A *static slice* of a program $P$ with respect to a slicing criterion $< s, v >$ is the set of all the statements of the program $P$ that might affect the slicing criterion for any possible inputs to the program. Korel and Laski [14] introduced the concept of a *dynamic program slicing*. A *dynamic slice* contains only those statements that actually affect the slicing criterion for a given execution. Dynamic slicing [3, 14] is used to find the slice of a program with a certain input. This approach results in a possible smaller slice, because some options in the control flow might be eliminated. Program slicing has been found to be useful in a variety of applications such as debugging, program understanding, maintenance, testing, model checking and program comprehension.

Aspect-Oriented Programming (AOP) [1, 2, 8, 10, 15, 17] is an emerging programming paradigm that was first published in June 1997 by Gregor Kiczale's team from Xerox Palo Alto Research Centre (PARC). AOP [1, 2, 15, 17] is a programming methodology which enables a clear separation of concerns in software applications. A concern in mean of software development can be understood as "a specific requirement that must be addressed in order to satisfy the overall system goal".

Many works have been done by several researchers on slicing of procedural and object-oriented programs [9, 11, 19, 20, 22, 23, 24, 32, 33]. But very few work have been carried out on slicing of Aspect-Oriented (AO) programs [21, 25, 27, 34, 35]. Though, aspects, pointcuts, join-points and advices are greatest strengths of AOP, they pose special difficulties in analyzing the AO programs. So, special care has to be taken. The existing slicing algorithms for procedural and object-oriented programs cannot be applied directly to AOP, due to the presence of some special features in AO programs. To the best of our knowledge, no work has been reported for dynamic slicing of concurrent AOPs.

In this paper, we propose an algorithm for dynamic slicing of concurrent AO programs. First, we develop an intermediate representation to represent a concurrent AO program. Then, we have extended the node marking dynamic slice (NMDS) algorithm proposed by Mohapatra, *et al.*, [18, 25] for computing the dynamic slices of concur-rent AOPs and named it as *Concurrent Aspect-oriented Dynamic Slicing (CADS) algorithm*. The rest of the paper is organized as follows: In Section 2, we present a brief introduction to concurrent AOP. In Section 3, we present the review of related work. In Section 4, we present some basic concepts and definitions that are to be used in dynamic slicing algorithm. Section 5, discusses the Concurrent Aspect-Oriented System Dependence Graph (CASDG), Intermediate Program Representation for representing concurrent AOPs. In Section 6, we describe the Concurrent Aspect-oriented dynamic slicing (CADS) algorithm. Section 7, discusses the implementation of CADS algorithm. Section 8 concludes the paper.

## 2. Concurrent AOP

In this section, we first discuss the basic concepts of concurrent AOP. Then, we discuss about AspectJ, a popular AOP language. Next we discuss some features of AspectJ.

### 2.1. Basic Concepts of Concurrent AOP

The scientist at the Palo Alto developed the first AO language called AspectJ, [6, 12] as an extension to Java language [26]. AOP [1, 2, 5, 8, 13, 17] gives the possibility to implement individual concerns in a loosely coupled way, and to combine these implementations to form the final system. There are two main properties of any AOP language [1, 2, 5, 13, 17] distinguishing it from other paradigms:

a) Quantification: It is the idea that a single statement or a number of statements may affect many statements in the program.

b) Obliviousness: It means that one cannot tell that the aspect code will execute only by examining its content.

Concurrency related concerns do not align well with class decomposition, and therefore source code related to such concerns suffers from the well-known negative phenomena of code scattering and tangling. Examples of concurrency-related concerns are: specification of tasks that can run in separate threads, definition of sections of behavior that must be subjected to synchronized access in order to avoid race conditions.

```
NON-ASPECT CODE                              ASPECT CODE
1. import java.util.*;                       61. public aspect ThreadAspect {
2. import java.util.Scanner;                 62. public pointcut classic(): execution(void run());
3. class Thread1 extends Thread {            63. before():classic(){
4. String name;                              64. System.out.println("Starting of aspect code for Thread1"); }
5. Thread t;                                 65. after():classic(){
6. Thread1(String Threadname){               66. System.out.println("Ending of aspect code for Thread1"); }
7. name=Threadname;                          67. public pointcut classical():execution(int rev());
8. t=new Thread(this,name);                  68. before():classical(){
9. System.out.println("1st Thread:"+t);      69. System.out.println("Aspect Code Starts for Thread2"); }
10. t.start();                               70. after()returning(int b):classical(){
11. meth1();                                 71. if(b%2== 0){
12. try{                                     72. System.out.println("The no is even:"+b+"\n"); }
13. for(int k=0; k<5;k++){                   73. else {
14. t.sleep(3000);                           74. System.out.println("The no is odd:"+b+"\n"); }
15. System.out.println("name = "+t +k);} }   75. System.out.println("Process Ends Completely"); } }
16. catch(InterruptedException e){ } }
17. public void run( ){
18. int i;
19. int j=200;
20. Scanner sc;
21. sc = new Scanner(System.in);
22. System.out.print("Enter value for i");
23. i = sc.nextInt( );
24. while(++i < --j);
25. System.out.println("The midpoint is :" + i); }
26. void meth1(){
27. int x=10;
28. int y=20;
29. System.out.println("the area is :" +x*y); } }
30. class Thread2 extends Thread{
31. String name;
32. Thread t;
33. Thread2(String Threadname){
34. name=Threadname;
35. t=new Thread(this,name);
36. System.out.println("2nd Thread:" +t);
37. t.start();
38. rev();
39. meth2(); }
40. int rev(){
41. int a,b=0;
42. Scanner sc;
43. sc = new Scanner(System.in);
44. System.out.println("Enter value for n:");
45. int n = sc.nextInt();
46. while(n>0) {
47. a=n%10;
48. b=a+b*10;
49. n=n/10; }
50. System.out.println("The result is :" +b+"\n");
51. return b; }
52. void meth2(){
53. int z=12;
54. int x=10;
55. int y=20;
56. System.out.println("The volume is:"+x*y*z+"\n"); } }
57. class ConcurrentThread{
58. public static void main(String args[ ]){
59. new Thread1("One");
60. new Thread2("Two");  }
}
```

**Figure 1. A Concurrent AspectJ Program**

Figure 1 shows a sample concurrent AspectJ program. In this program, there are two threads, named Thread1 and Thread2. Each one extends the Thread class. The first thread, Thread1 prints value from 0 to 4 with a sleeping interval of 3000 ms inside a try-catch block. Then, it computes the mid-point of two numbers. It also calculates the area of a square. The second thread, Thread2 calculates the reverse of a number and then calculates the volume of a cube. Both the Threads are having constructors, inside which the thread get a start.

### 2.2. AspectJ – A popular AOP language

The most popular AOP language is *AspectJ*. *AspectJ* was created by Chris Maeda at Xerox Palo Alto Research Center (PARC). This is a general-purpose, AO *extension* to Java programming language. The *AspectJ compiler (ajc)*, [15] combines Java and aspect source files as well as Java Archive (JAR) files to create woven class files or Java Archive (JAR) files as output. Java Archive (JAR) files aggregates many files into one file and is used to distribute Java applications and libraries. The weaving process in AspectJ is done during compilation process. The base source code and the code defining weaving rules are processed to create an output class file or a Java Archive (JAR) file. Figure 2 shows the AspectJ compilation process.

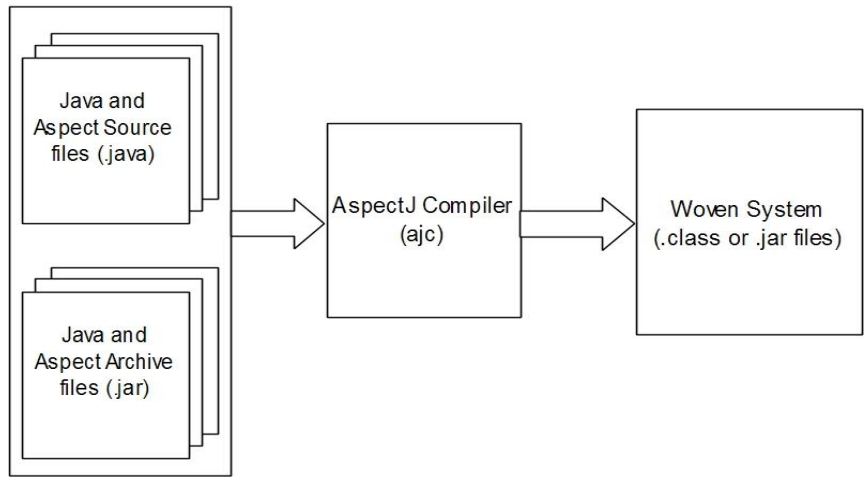

**Figure 2. AspectJ Compilation Process**

### 2.3. Features of AspectJ

AspectJ [1, 2, 5, 8, 10, 12, 17] is not the substitute for Java [26], rather it's an extension to Java language. It adds certain features to Java language. These features include Join Points, Pointcuts, Advice, Aspect, Introductions etc. We explain these features below.

**Join Points:** These are the well defined points in the execution of programs, such as method call, method execution, and method reception. *Join point* is a fundamental concept of AspectJ identifying an execution point in a system.

**Pointcuts:** *Pointcut* is a program construct used to select the join points as well as the collecting context at those joins points. A *pointcut* may select a call to a method and capture the method's context. In other words, we may say that *pointcuts* specifies the weaving rules and join point represent the situations satisfying those rules.

**Aspect:** *Aspects* in AOP, is similar to *classes* in OOP. *Aspect* describes how pointcuts and advices should be combined together. *Aspects* include poincuts, advices, and introductions which are used to add a public or private method, field, or interface implementation declaration into a class.

**Advice:** *Advice* is the code executed at a join point. *Advices* are selected by pointcuts. It is possible to execute advices before, after, and around a join point. The *advice* code looks *similar* to a *method*. There are three types of advices in AspectJ. They are: before, after and around.

1. *before - Before advice* runs as a join point is reached before the program proceeds with the join point.

2. *around* - The *around advice* on a join point runs as the join point is reached, and has explicit control over whether the program proceeds with the join point.

3. *after* - The *after advice* on a particular join point runs after the program proceeds with that join point.

   Additionally there are *two* special cases of *after advice: after returning and after throwing*, corresponding to the two ways a sub-computation can return through a join point.

   a) *after returning - after returning advice* runs just after each join point picked out by the pointcut, but only if it returns normally. The return value can be accessed. After the advice runs, the return value is returned.

   b) *after throwing - after throwing advice* runs just after each join point picked out by the pointcut, but only when it throws an exception. The advice re-raises the exception after it is done.

**Introduction:** An *introduction* can be used to make static changes to application modules such as adding methods or fields to a class. It allows an aspect to add methods, fields or interfaces to existing classes.

## 2.4. Concurrency Model of AspectJ

*Concurrency* is an important factor in the behavior and performance of modern code. Concurrent programs are difficult to design, write, reason about, debug, and tune. AOP approaches typically provide a translation mechanism that "weaves" aspects into base programs. In a concurrent environment, this would correspond to a monolithic weaving strategy yielding a concurrently executing program. AOP is concerned with the modularization of so-called *crosscutting functionalities*, which cannot be reasonably modularized using traditional programming means, such as objects and com-ponents. Concurrency can be introduced and controlled by exploiting existing libraries for concurrent programming, e.g., Java's thread library. Currently there exist no AOP models with facilities for the definition of concurrently executing aspects as well as for the coordi- nation of aspects and concurrent base programs directly in terms of AOP specific concepts.
Aspects should provide the following support for concurrent executions.

1. Aspects are most frequently defined through matching of "pointcuts" on base program executions and executing "advice(s)" when a match occurs, thus modifying the base program execution. Stateful aspects, allows the explicit representation of relations between execution events, and very useful for coordinating aspects and base programs, and provide information crucial for the verification of AO programs. Stateful aspects are only supported in AspectJ through *cflow pointcuts*. Concurrent aspects should therefore support stateful pointcuts and aspects.

2. In case of multiple concurrently executing aspects of multiple advices at one point during base program execution, program may trigger the execution of another advice instead of the base action. A *concurrent AOP* should allow partial orderings on the advices, which defines the parts for concurrent execution and the parts to be synchronized later on.

The use of concurrency in AO programs has mainly two advantages. First, traditional support to concurrency features degrades performance, even when the features are not used, e.g., in Java, each object implements a monitor even when concurrency features are not used in the program. A similar benefit was observed in middleware systems where, performance improvement was attained by extracting non-core features to aspects. Second, treating concurrency issues as additional features modeled by aspects helps to develop concurrent applications, where concurrency issues are more *modular* and can be *unplugged*. The aspect must manage history of each join point context, maintaining data structures for each join point context. This can be performed through a collection that associates each join point context to particular data structures. Moving the core concern code to an aspect code introduces overheads related to aspect instantiation and management, additional objects and method calls.

## 3. Review of Related Work

Many works have been reported on slicing of procedural and object-oriented programs [9, 11, 19, 20, 22, 23, 24, 32, 33]. But, few works have been reported on slicing of AOPs [21, 25, 27, 34, 35]. Zhao [34] was first to develop the *Aspect-oriented System Dependence Graph* (ASDG) to represent AO programs. The ASDG is constructed by combining the SDG [16] of non-aspect code, the aspect dependence graph (ADG) of aspect code and some additional dependence arcs used to connect the SDG and ADG. Then, Zhao used the two-phase slicing algorithm [16] proposed by Larsen and Harrold [16] to compute static slice of AO programs. The main limitation of this approach [34] is that, Zhao has not addressed the concepts of pointcuts clearly.

SDG construction algorithm proposed by Zhao and Rinard [35] provides an efficient way for addressing the concepts of pointcuts. The SDG construction algorithm [35] is concrete in nature and can be used for constructing SDG [11, 16] of an AO program.

Timor ter Braak [27] extended the ASDG proposed by Zhao [34, 35] to include inter-type declarations or introductions in the intermediate dependency graph. Each inter-type declaration was represented in the form of a field or a method as a successor of the particular class. Then, Braak [27] used the two-phase slicing algorithm of Horwitz, *et al.*, [11], to find the static slice of an AO program. Thus the limitations of the SDG construction algorithm proposed by Zhao and Rinard [35] *i.e.* addressing the concept of weaving properly gets resolved quite efficiently as it includes the introductions in the intermediate representation.

All the above mentioned techniques [27, 34, 35] provide a mean for the computation of static slices of an AOP. Mohapatra, *et al.*, [21] introduced the methodology for computing dynamic slice of an AOP. They presented the Trace Based Dynamic Slicing (TBDS) algorithm [21] for computing the dynamic slice of AOPs. They have used Dynamic Aspect-oriented Dependency Graph (DADG) [21] as the intermediate representation. The limitation of this approach [21] is that, it stores each occurrence of a statement in an execution trace which will take a lot of spaces and time. Thus, it yields more time and space complexities.

Mohapatra, *et al.*, [25] also proposed a node marking dynamic slicing (NMDS) algorithm for computing the dynamic slice of an AOP. This algorithm is a quite efficient dynamic slicing technique, compared to the TBDS algorithm [21]. This is because, in this algorithm Extended Aspect-oriented System Dependence Graph (EASDG) is used to represent the aspect oriented software, and no trace file is used for storing the execution history. Also, there is no need for creating any other nodes during the run time.

To the best of our knowledge, no works have been reported for dynamic slicing of concurrent AOPs. In this paper, we propose an approach for computing dynamic slices of a concurrent AOP.

## 4. Some Basic Concepts and Definitions

In this section, we present some basic definitions, concepts and terminologies associated with the intermediate program representation. These concepts and definitions are later used throughout the discussion in this paper. The concepts and definitions of Def(var), Use(var), DefSet(var) and UseSet(var) are available in [22, 23]. Also the basic concepts of communication dependence and synchronization dependence are available in [18]. However, we present them here for the *sake of completeness*. Further, we introduce some new concepts, definitions and terminologies which will be used in the rest of the paper.

### 4.1. Definitions

**Definition 1 (Precise Dynamic Slice):** A dynamic slice is said to be *precise*, if it contains only those statements that *affect the value of the variable* at a program point for execution.

**Definition 2 (Correct Slice):** A *correct slice* contains all the statements that *affect* the slicing criterion.

**Definition 3 (Def(var)):** Let *var* be a variable in a class in the program *P*. A node *v* of the CASDG is said to be a *def(var) node,* if *v* represents a definition(assignment) statement that defines the variable *var*. In the CASDG of Figure 5, node 53 is the *Def(z)* node.

**Definition 4 (DefSet(var)):** The *set DefSet(var)* denotes the set of all *def(var)* nodes. In the CASDG of Figure 5, the DefSet(z)={53}.

**Definition 5 (recentDef(var)):** For each variable *var*, *recentDef(var)* represents the node corresponding to the *most recent definition* of *var* with respect to some point *s* in an execution.

**Definition 6 (Use(var)):** Let *var* be a variable in a program *P*. A node *v* of the CASDG of *P* is said to be a *Use(var) node*, if *v* represents a statement that *uses the value* of the variable *var*. In the CASDG of Figure 5, node 56 is theUse(z) node.

**Definition 7 (UseSet(var)):** The *set UseSet(var)* denotes the set of all Use(var) nodes. In the CASDG of Fig. 5, UseSet(z)={56}.

**Definition 8 (Control Dependence):** *Control Dependences* represents the *control flow relationships* of a program i.e. the predicates on which a statement or an expression *depends* during execution.

**Definition 9 (Data Dependence):** *Data Dependences* represent the relevant *data flow relationships* of a program i.e. the flow of data between statements and expressions. Let *C* be a CASDG. Let *m* be a *Def(var)* node and *n* be a *Use(var)* node. The node *n* is said to be data-dependent on node *m,* if there exists a directed path *Q* from *m* to *n* such that there is no intervening *Def(var)* node in *Q*.

**Definition 10 (Thread level Dependence):** The dependencies that are associated among threads are termed as Thread level Dependence. This type of dependency can arise among a single thread, as well as between multiple threads.

**Dependencies among multiple threads:** This type of dependency is termed as *Synchronization Dependence* as it includes *Synchronization* operations. A statement *m* in one thread is said to be synchronization dependent on statement *n* in another thread iff, the execution of *m* is *dependent on execution* of *n* due to a *synchronization operation*. Java language provides methods like *wait(), notify() and notifyall()* for achieving synchronization.

**Dependencies among single thread:** This type of dependency arises within a single thread. Let *y* be a sleep node and *x* be a statement associated with the sleep node. Statement *y* is said to be Non-synchronized Dependent, iff, the value of statement *y* has an impact on statement *x*. That is the value defined at statement *y* depends on *x* for execution. This type of dependency is termed as *Non-synchronized* because it *doesn't involve* any *synchronization* operations. Java provides methods like sleep(), suspend(), spawn() etc. that are used to show dependency between single threads. For example, in Figure 3, in Thread 2, statement 14 is dependent on statement 15 and statement 13 for execution which doesn't involves any synchronization operations. So, there exists a *non-synchronization dependency*, between statement 14 and 15.

**Definition 11 (Exception Check Dependence):** This type of dependency arises when the control of the program goes inside the try-catch block. This type of dependency is purely dependent on the execution of the program. If the control goes inside the try-catch block then it will first execute the try block statements, and then its corresponding catch block statements. For a SDG, of non-aspect code, let *x* be the node representing try statement of the try-catch block. A node *y* is said to be *exception check dependent* on statement *x*, iff there exists a directed path (DP) from *x* to *y*, such that none of the nodes in DP constitutes a try statements.

**Definition 12 (Communication Dependence):** Informally, a statement *y* in one thread is said to be *communication dependent* on statement *x* in another thread, if the value of *x* is directly used at *y* through *inter thread communication*. For example in Figure 3, node 28 of Thread1 is communication dependent on node 55 of Thread2. Similarly node 27 of Thread1 is communication dependent on node 54 of Thread2. In the rest of paper, we use the terms nodes and vertices interchangeably.

## 5. Intermediate Program Representation

This section, discusses definitions as well as the construction of concurrent aspect-oriented system dependence graph (CASDG).

### 5.1. Concurrent Aspect-oriented System Dependence Graph (CASDG)

In this section, we discuss the definition and construction of CASDG in the context of concurrent AO programs. Our CASDG is an extension to the ASDG of Zhao, *et al.*, [34, 35]. The CASDG of a concurrent AO program consists of three parts: a *System Dependence Graph (SDG) for non-aspect code*, *a group of dependence graphs for aspect code* and *some additional dependence edges for connecting the system dependence graph for non-aspect code to the dependence graphs for aspect code*. The CASDG is modelled by the main program together with all the methods, constructors in the non-aspect code and all the pointcuts, advices and introductions in the aspect code. A concurrent AspectJ program is divided into two parts: *base code or non-aspect code,* includes classes, threads, interfaces and other standard Java constructs and *aspect code,* implements the *cross cutting concerns* in the program. For example, Figure 1 shows a concurrent AspectJ program. In this example program, there are two thread classes each extending from Thread class. In the base or non-aspect code, Class ConcurrentThread consists of standard simple Java constructs such as constructors, inheritance, exception, Threads and the aspect code, ThreadAspect contains the pointcut and advices. Any implementation of AspectJ ensures that both the codes i.e. aspect code and non-aspect code run together in a properly and coordinated manner. Such a process is termed as *aspect weaving*. The key component for this process is the *aspect-weaver* which makes the applicable advices to run at the appropriate join points.

**5.2. Construction of CASDG**

The construction of CASDG consists of the following steps:

- Construction of SDG for non-aspect code of the program.
- Construction of Aspect Dependence Graphs (ADGs) for the aspect code of the program.
- Determination of weaving points and inserting the weaving vertices into the SDG of the non-aspect code.
- Weaving the system dependence graph (SDG) and the aspect dependence graph (ADG) at weaving vertices to complete the CASDG by adding some special kinds of dependence edges between the SDG and the ADG.

**5.2.1. Construction of SDG for non-aspect code:** The system dependence graph (SDG) [11, 16] of an object-oriented program (OOP) or the non-aspect code of an AO program is a collection of *method dependence graphs (MDGs)*. A *method dependence graph (MDG)* for a method contains *vertices* representing *statements or predicates* of the methods and *edges* representing *data and control dependences* among the statements. Each MDG has a unique vertex called method start vertex to represent the entry of the method. Figure 3 shows the SDG of Figure 1.

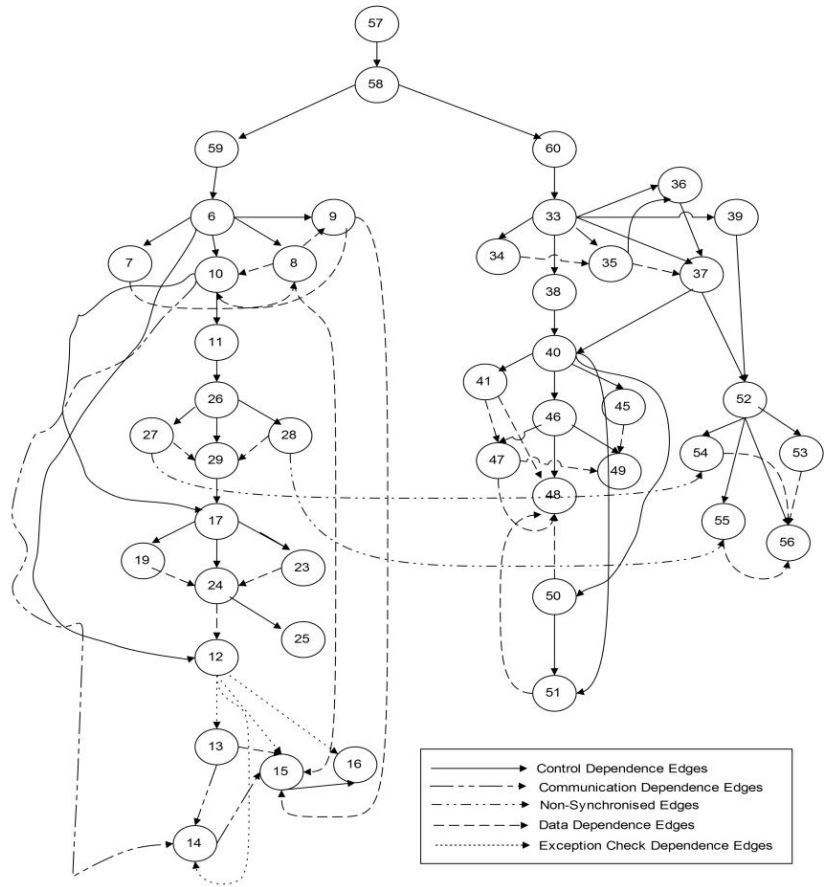

**Figure 3. System Dependence Graph of the Sample Program Given in Figure 1**

To model *parameter passing*, there is *formal-in vertex* for each formal *parameter* of the method and a *formal-out vertex* for each *formal parameter* that may be *modified* by the method. The SDG also contains *call vertex* and *actual parameter vertices* to associate with each call site. At each call vertex, there is an *actual-in vertex* for each *actual parameter* and an *actual-out vertex* for each *actual parameter* that may be *modified* by the called method. The construction of the complete SDG can perform by connecting all the MDGs at call sites. A *call arc* represents the *call relationships* is added between the call vertex of the calling method's MDG and the start vertex of the called method's MDG. *Actual-in and formal-in vertices* are connected by *parameter-in arcs* and *formal-out and actual-out vertices* and are connected by *parameter-out arcs*. These *parameter arcs* represent the parameter passing. The *summary arcs* are used to represent the *transitive flow* of dependencies in the SDG, and are created by *connecting* an *actual-in vertex* to an *actual-out vertex*, if the value associated with the *actual-in vertex affects* the value associated with the *actual-out vertex*.

**5.2.2. Construction of dependence graphs for aspect code:** The dependence graph for aspect code, called the *Aspect Dependence Graph (ADG)*, is constructed by combining the *advice dependence graph, introduction dependence graph, pointcut dependence graph and method dependence graph*. *Advice dependence graph* is used to represent *advice* in an aspect. The *ADG* of *advice* is similar to the *MDG* of a *method* such that its *vertices* represent *statements or predicate* expressions in the advice, and its *arcs* represent *control or data dependencies* between vertices. *Introduction dependence graph (IDG)* is used represent an *introduction* in an aspect. The *IDG* of an *introduction* is similar to the *MDG* of method such that its vertices represent statement or a control predicate of a conditional branch statement in the introduction and its arcs represent control or data dependencies between these statements. *Pointcut dependence graph* is used to represent a *pointcut* in an aspect. As Pointcuts declared in an aspect contain no body, so for each pointcut designator, the pointcut start vertex need to be constructed and that will be the entry point to the pointcut. *Method dependence graph* is used to represent a method and associated dependencies in an aspect.

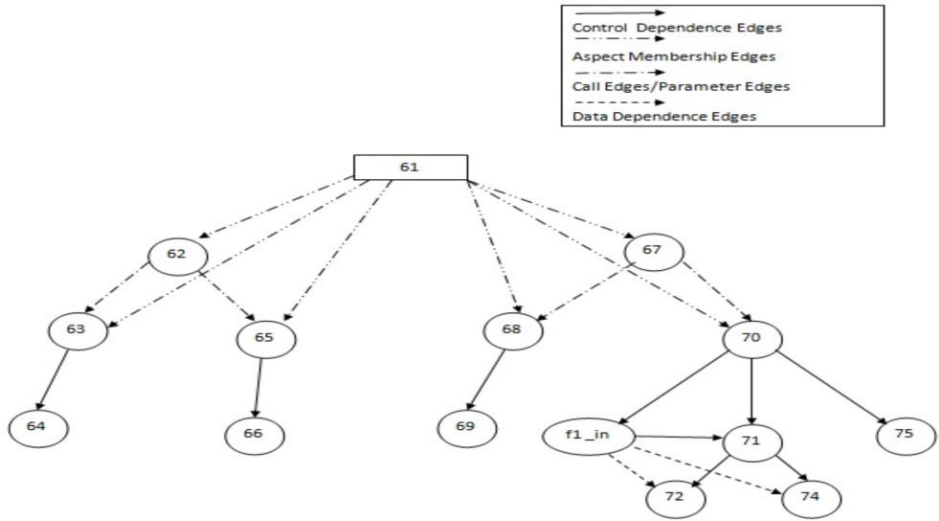

**Figure 4. ADG of the Aspect Code given in Figure 1**

The graph for advice, introduction, pointcut and method in an aspect code is similar to the method dependence graph of base code. The graph for each piece of advice, introduction, pointcut and method are connected using call edges, parameter edges, data dependence edges. There is a unique vertex called *aspect entry vertex* in *ADG* which represents the entry into an aspect. The various members of an aspect such as advice, introduction, pointcut and method are connected to the *aspect entry vertex* through *aspect membership edges*. Figure 4 shows ADG of Figure 1.

**5.2.3. Determination of weaving points:** The weaving points are determined from the non-aspect code with the help of *join-points*. The *join-points* are well defined points along the execution of a program. These can be: a method execution or call, read/write access to a field, the instantiation of an object or class and throwing an exception. In AspectJ, *join points* are defined in each aspect of an AO program with the *pointcut designator*. *Pointcuts* are further used in the definition of *advice*. By careful examination of join points, declared in the pointcuts and their associated advice, the weaving vertices can be determined to connect the SDG of the non-aspect code to the ADG of the aspect code.

**5.2.4. Weaving the SDG and the ADG:** To construct the complete CASDG for an AspectJ program, the SDG of the non-aspect code and the ADG of the aspect code at *weaving vertices* are connected by using *weaving edge*.

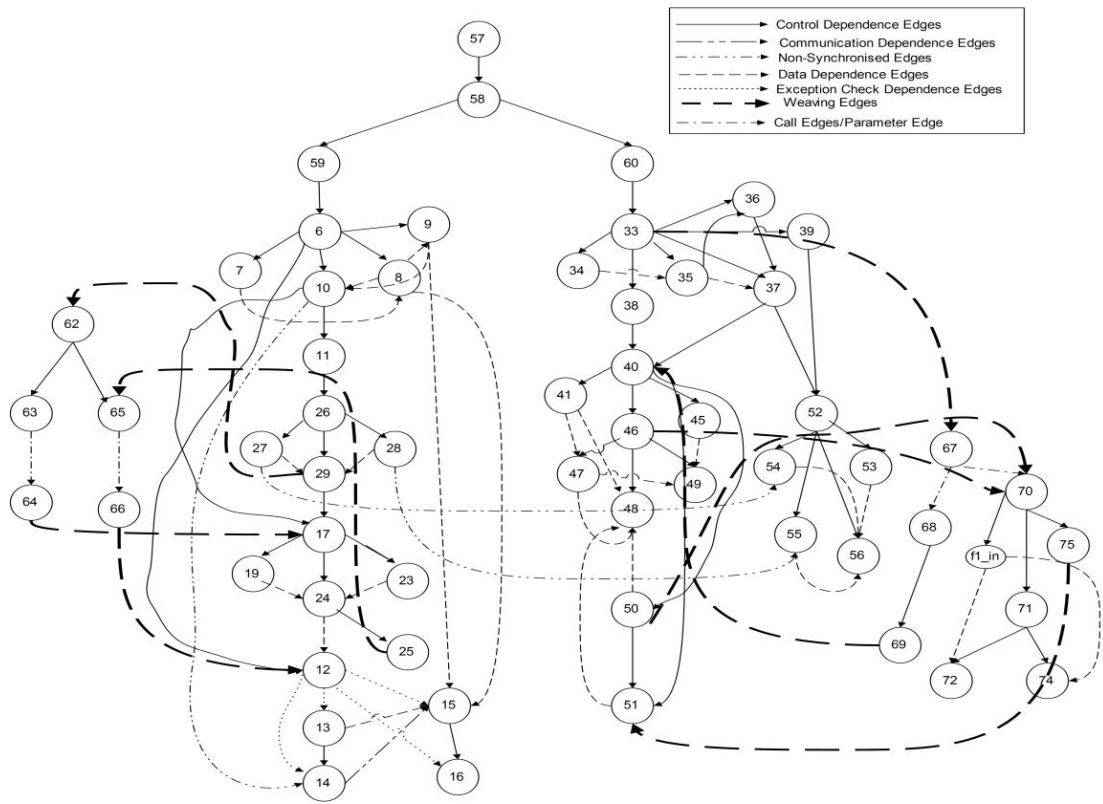

**Figure 5. CASDG of the Sample Program Given in Figure 1**

A *call arc* is added between a call vertex and the start vertex of the ADG, IDG, or MDG of the called advice, introduction or method. Actual and formal parameter vertices are connected by *parameter arcs*. A complete CASDG is shown in Figure 5.

## 6. Concurrent Aspect-oriented Dynamic Slicing (CADS) Algorithm

Before the execution of a concurrent AO program *P*, its Concurrent Aspect Oriented System Dependence Graph is constructed statically only once. During the execution of the program *P*, the algorithm marks and unmarks the executed nodes appropriately. To handle the method/constructor calls, when a statement invokes a method, the algorithm marks the corresponding call vertex and the associated actual parameter vertices. Simultaneously the algorithm marks the method entry vertex of the corresponding called methods and the associated formal parameter vertices. After executing a method call, and recording its dynamic slice at the invoking node, all the marked nodes associated with the method entry vertex are unmarked. Let dynamicSlice(w,var) denote the dynamic slice with respect to the variable *var* for the most recent execution of the statement *s*, corresponding to node *w*. Let $x_1, x_2, x_3,..., x_k$ be all the marked predecessor nodes of *w* in the CASDG after an execution of the statement corresponding to node *w*. Then the dynamic slice with respect to variable *var* for the present execution of the statement *s* corresponding to node *w* is given by:

dynamicSlice(*w*, *var*) = [$x_1, x_2, x_3,..., x_k$] ∪ dynamicSlice($x_1$,*var*) ∪ dynamicSlice($x_2$,*var*) ∪ dynamicSlice($x_3$,*var*) ∪,...,∪ dynamicSlice($x_k$, *var*)          Equation (1).

The CADS algorithm computes the dynamic slice with respect to the slicing criterion simply looking up the corresponding dynamic slice, computed during run-time. The CADS algorithm can be applied to an AO program after constructing the corresponding CASDG. The pseudo code of the algorithm is given below in the next subsection.

### 6.1. Algorithm: Concurrent Aspect-oriented Dynamic Slicing (CADS) Algorithm

**Step 1. Initialisation:**

(a) Unmarks all the nodes of the CASDG.

(b) Set dynamicSlice(*w*,*var*) = Ø for every *var* of each node *w* of the CASDG.

(c) Set recentDef(var) = NULL for every variable *var* of the program *P*.

// End of Initialisation.

**Step 2. Run-time Updations:** At run-time until a program ends or a slicing criterion is given, carry out the following after each execution of each statement *s* of the program *P*. Let node *w* in CASDG corresponds to the statement *s* in program *P*.

(a) for every variable *var* used at node *w*, update Equation (1).

(b) If w is a Def(var) node, then do the following:

    i. update dynamicSlice(w,var) = w ∪ dynamicSlice(w,var).

    ii. unmark the node recDef(var).

    iii. update recDef(var) = *w*.

(c) Mark the node *w*.

(d) If w is a call vertex, then do the following:

  i. mark the vertex *w*.

  ii. mark the actual-in and actual-out vertices associated with *w* corresponding to the present execution of *w*.

  iii. mark the method entry vertex of the corresponding called method for the present execution of the vertex *w*.

  iv. mark the formal-in and formal-out vertices associated with the method entry vertex.

(e) If *w* is a pointcut start vertex, then do the following:

  i. mark the vertex *w*.

  ii. mark the actual-in and actual-out vertices associated with *w* corresponding to the present execution of *w*.

  iii. mark the method corresponding advice start vertices for the present execution of the vertex *w*.

  iv. mark the formal-in and formal-out vertices associated with the advice start vertices of the pointcut.

(f) If *w* is a weaving vertex, then do the following:

  i. mark the vertex *w*.

  ii. mark the vertex present on another side of weaving edge corresponding to the present execution of *w*.

(g) If *w* is a vertex representing sleep () method of a thread, then do the following:

  i. mark the vertex *w*.

  ii. mark the vertex associated with the sleeping edge corresponding to the present execution of *w*.

(h) If *w* is a vertex representing try statement of a try-catch block, then do the following:

  i. mark the vertex *w*.

  ii. mark the actual-in and actual-out vertices associated with *w* corresponding to the present execution of *w*.

  iii. mark the method corresponding try-catch block statements for the present execution of the vertex *w*.

  iv. mark the formal-in and formal-out vertices associated with the try-catch block statements of the Exception block.

// end of Run-Time updations.

**Step 3. Slice Look-Up:**

(a) If a slice command <s,var> is given then carry the following:

    i.   look up dynamicSlice(w,var) for variable var for the content of slice.

    ii.  display the result.

(b) if program has not terminated, then go to step 2.

### 6.2. Working of CADS Algorithm

We illustrate the working of our algorithm with the help of an example. Consider the sample Concurrent AspectJ program given in Figure 1. The SDG corresponding to the non-aspect code or base code is given in Figure 3. The ADG corresponding to the aspect code is shown in Figure 4. The complete CASDG is shown in Figure 5. During the initialization step, our CADS algorithm first unmarks all the nodes of the CASDG and sets dynamicSlice(w,var) = Ø, for every node w of the CASDG.

Now, for the input data $i = 123$ and $n = 365$, the program executes the statements 58, 59, 6, 7, 8, 9, 10, 11, 26, 27, 28, 29, 62, 63, 64, 17, 19, 23, 24, 25, 65, 66, 12, 13, 14, 15, 13, 14, 15, 13, 14, 15, 13, 14, 15, 13, 14, 15, 60, 33, 34, 35, 36, 37, 38, 39, 67, 68, 69, 40, 41, 45, 46, 47, 48, 49, 51, 50, 70, 71, 72, 74, 75, 52, 53, 54, 55, 56 in order. So, our CADS algorithm marks the vertices 58, 59, 6, 7, 8, 9, 10, 11, 26, 27, 28, 29, 62, 63, 64, 17, 19, 23, 24, 25, 65, 66, 12, 13, 14, 15, 60, 33, 34, 35, 36, 37, 38, 39, 67, 68, 69, 40, 41, 45, 46, 47, 48, 49, 50, 51, 70, 71, 72, 74, 75, 52, 53, 54, 55, 56. The algorithm also marks the associated actual parameter vertices at call site and the formal parameter vertices at the called method. Suppose, we want to compute the dynamic slice with respect to slicing criterion <56, z>. Now, while computing the dynamic slice with respect to variable z, according to CADS algorithm the dynamic slice with respect to variable z at statement 56 is given by:

dynamicSlice(56, z) = {52,53,54,55} ∪ dynamicSlice(52, z) ∪ dynamicSlice(53, z) ∪

dynamicSlice(54, z) ∪ dynamicSlice(55, z).

By evaluating the expression in a recursive manner the final dynamic slice is obtained. The slice will contain the statements 56, 53, 54, 52, 55, 27, 39, 37, 28, 26, 33, 35, 36, 11, 60, 34, 10, 58, 57, 6, 9, 8, 59, 7. Now, we shall find the backward dynamic slice computed with respect to variable i at statement 25 of Thread1's run() method. According to the CADS algorithm, the dynamic slice with respect to variable i at statement 25 is given by:

dynamicSlice(25,i) = {24} ∪ dynamicSlice(24,i).

By evaluating the expression recursively, the final dynamic slice is obtained. The slice will contain the statements 25, 24, 23, 17, 19, 29, 64, 10, 26, 27, 28, 63, 8, 6, 9, 11, 62, 7, 59, 58, 57. The statements included in the dynamic slice are shown as shaded vertices in Figure 6.

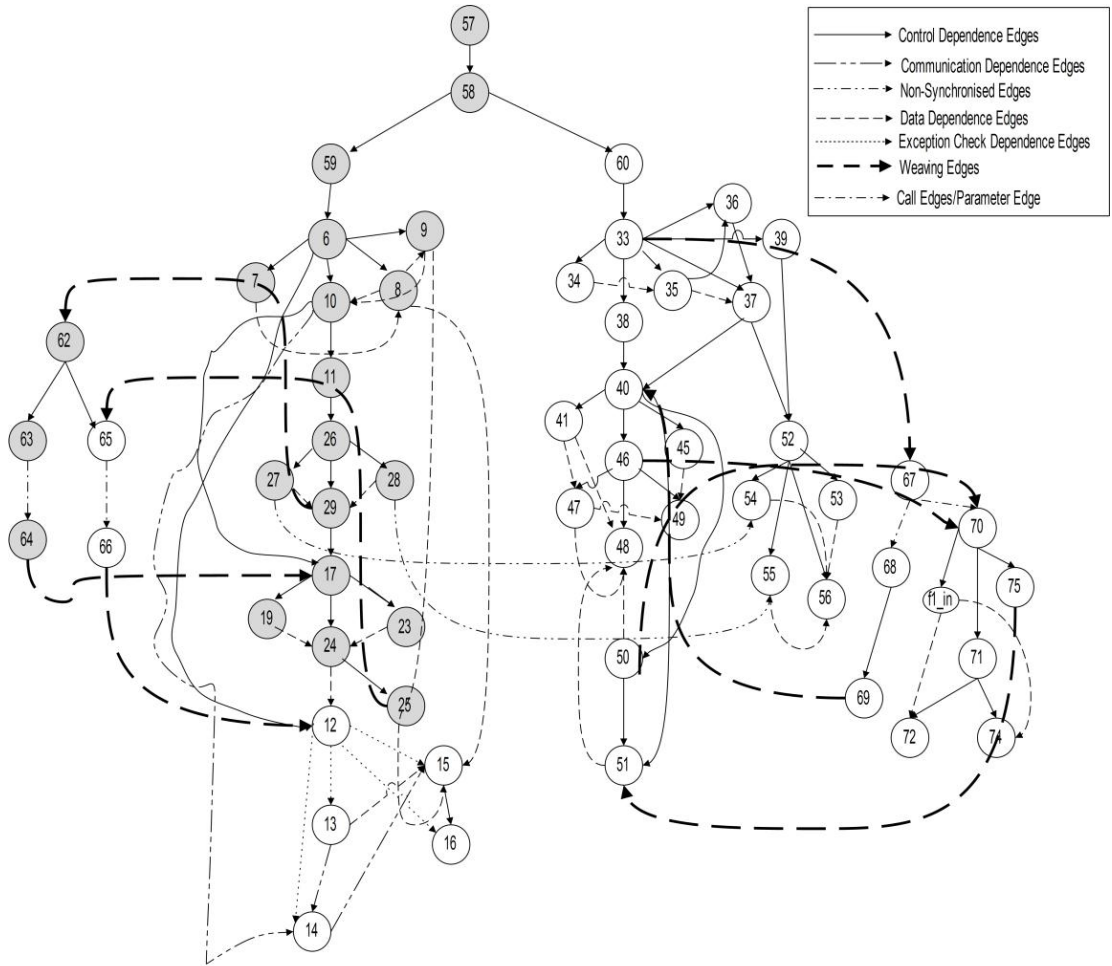

**Figure 6. The Updated CASDG of the Sample Program Given in Figure 1**

**6.3. Complexity Analysis**

In this subsection, we analyze the space and time complexities of our CADS algorithm.

*Space Complexity:* Let *P* be a concurrent AO program having *n* statements. The constructed CASDG are directed graphs on *n* nodes. Note that a graph on *n* nodes with optionally marked edges requires $O(n^2)$ space. So, the space requirement for CASDG of *P* with optionally marked edges is $O(n^2)$. We need the following additional run-time space requirement for computing the intermediate program representation:

1. To store dynamicSlice(*w,var*) for every node *w* of CASDG at most $O(n^2)$ space is required.

2. To store recentDef(*var*) for every variable var of *P*, at most $O(n)$ space is required. So the *space complexity* of the *CADS algorithm* is $O(n^2)$, where *n* is the total number of executable statements of program *P* or number of nodes in the CASDG.

*Time Complexity:* To determine the *time complexity* of the CADS algorithm, we need to consider two factors. The *first* one is the *execution time requirement* for the run-time updations in CASDG. The *second* one is the *time required to look up* the data structure dynamicSlice for *extracting* the dynamic slice, once the slicing command is given. To determine the *run-time complexity*, we need to find the time required for updating the relevant information and computing the dynamic slice after execution of a statement in the program. After execution of a statement of the program (node *w* of the CASDG), we need to compute Equation (1). Now, let us calculate the maximum number of nodes that can be adjacent to *w* in the updated CASDG after an execution of the node *w* in an actual run of the program. Let the number of variables used at *w* be *m*. Then, the maximum number of nodes that can be adjacent to *w* is *m+6*. The additional vertices are meant for *control dependence, communication dependence, non-synchronized dependence, exception check dependence, weaving and call vertices*. Let the value of this number *m+6* be *s*. Now, we have to compute the time requirement for finding the union of all *s* sets, each sets having at most $O(n^2)$ elements, where *n* is the number of nodes in the CASDG or equivalently the number of executable statements in program *P*. The complexity of the set union is known to be $O(n^2s)$. So, the *worst case time complexity* of our algorithm is $O(n^2s)$.

### 6.4. Correctness of CADS Algorithm

In this subsection, we sketch the proof of correctness of our ENMDS algorithm.

Theorem: ENMDS algorithm always finds a correct dynamic slice with respect to a given slicing condition.

Proof: We can prove this through mathematical induction. Let *P* be any given concurrent AO program for which a dynamic slice is to be computed by using CADS algorithm. According to definition, for any set of input values to the program, the dynamic slice with respect to the first executed statement is certainly correct. Using this argument, we establish that the dynamic slice with respect to the second executed statement is also correct.

During execution of the program *P*, assume that CADS algorithm has produced correct dynamic slices prior to the present execution of a node *s*. Let *var* be a variable used at *s*, and dynamicSlice(*s,var*) be the dynamic slice with respect to the slicing criterion <*s,var*> for the present execution of the node *s*. Let node *d* = recentDef(var) is the reaching definition of the variable *var* for the present execution of the node *s*. Note that the node *d* is executed prior to the current execution of the node *s*. It is obvious that dynamicSlice(*d,var*) contains all the nodes which have affected the current value of the variable *var* used at *s*, since our CADS algorithm has marked all the incoming edges to *d* only from those nodes on which node *d* is dependent. If a node has not affected the variable *var*, then it will not be included in dynamicSlice(*d,var*). So, dynamicSlice(*d,var*) is a *correct* dynamic slice.

Further, the steps 2(d), 2(e), 2(f), 2(g), 2(h) of the algorithm ensure that the node *s* is dependent (with respect to its present execution) on node *v* if and only if the edge (*s,v*) is marked in the CASDG of the program *P*. Let $d_1, d_2, d_3,..., d_k$ be all the nodes on which *s* is dependent with respect to its present execution.

Then dynamicSlice(*s, var*) = [$d_1, d_2, d_3,..., d_k$] ∪ dynamicSlice($d_1$,*var*) ∪ dynamicSlice($d_2$,*var*) ∪ dynamicSlice($d_3$,*var*) ∪ ,..., ∪ dynamicSlice($d_k$, *var*).

Since dynamicSlice($d_1$,*var*)...dynamicSlice($d_k$,*var*) are all correct dynamic slices, the dynamic slice dynamicSlice(*s, var*) computed by step 2 of the algorithm must also be correct. Further step 3 of the algorithm guarantees that the algorithm stops when execution of the program *P* terminates. This establishes the correctness of the algorithm.

## 7. Implementation

In this section, we present an implementation of our CADS algorithm. We have named our dynamic slicing tool as dynamic slicer for concurrent aspect-oriented programs (DSCAP). In the following subsection, we briefly discuss the design of our tool.

### 7.1. Overview of the Design of DSCAP

The working of the slicing tool is schematically shown in Figure 7. The arrows in the figure show the data flow among the different blocks of the tool. The blocks shown in *rectangular boxes* represent *executable components* and the blocks shown in *ellipses* represent *passive components* of slicing tool. A program written in AspectJ is given as input to DSCAP. The overall control for the slicer (DSCAP) is done through a coordinator with the help of a graphical user interface (GUI). The coordinator takes user input from the GUI, interacts with other relevant components to extract the desired results and returns the output back to the GUI. The *ajc Preprocessor* reads the aspect headers, preprocesses the Java source code and determines where the given aspects must take effect. It then weaves the aspects into the code, generating new source files. After completion of weaving process, the control is passed to the *Lexical Analyzer*. The *Lexical Analyzer* reads the input program and breaks it into *tokens* for the grammar used in generating an LALR(1) Parser. The *lexical analyzer* component reads the concurrent AOP and breaks it into tokens for the grammar expressed in the parser. When the lexical analyzer component encounters a useful token in the program, it returns the token to the *parser* describing the type of encountered token. The *parser and semantic analyzer* component functions as a state machine. The parser takes the token given by the lexical analyzer and examines it using the grammatical rules laid for the input programs. The *semantic analyzer* component captures the following important information of the program. The information are: For each vertex $w$ of the program $P$, the semantic analyzer component captures the lexical successor and predecessor vertices of $w$ and type of variables defined and used at vertex $w$.

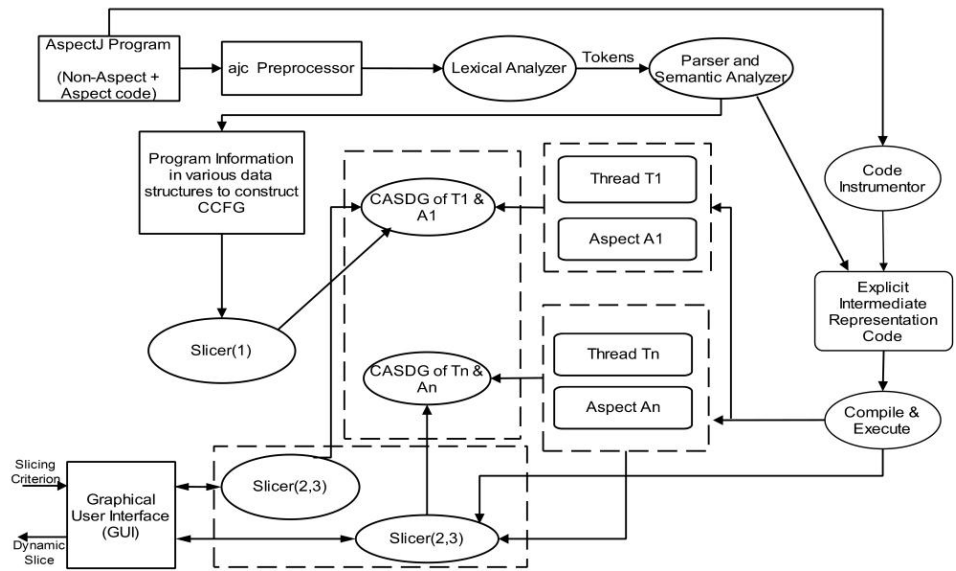

**Figure 7. Architectural Representation of the Working of the Slicer**

The basic information regarding objects, program, statement, variables and scope is captured in the CCFG of the program. The information available in the CCFG is used to construct the CASDG in Slicer(1) block. We have named this block as Slicer(1), as it corresponds to the step 1 of our CADS algorithm. The source code of the concurrent AO program is instrumented with the parametrized method calls to the slicer. This is achieved through the *Code-Instrumentor* block. The *Code-Instrumentor* block takes the concurrent AOP as input and attached calls to the slicer module after every executable statement. The *code-instrumentor* generates an explicit *intermediate representation code* corresponding to source program. This *explicit intermediate representation code* is then submitted to the *Compile and Execute Block*. The *Compile and Execute* block compiles, translates, links and execute the explicit intermediate representation code. The Slicer(2,3) blocks implements the step 2 and 3 of our CADS algorithm. That is why, we have given the name of this block as Slicer(2,3). The link between *Compile and Execute* block, Slicer(2,3) block and CASDG depicts how the explicit intermediate representation code calls the slicer, and how the slicer maintains the CASDG at run-time. The Slicer(2,3) block takes the slicing criterion as input from the Graphical User Interface (GUI) block and outputs the computed dynamic slice of concurrent AO program. The GUI block encapsulates the complexities involved in the functioning of all the other blocks by supporting user-friendly interface to the user.

## 8. Conclusion

We have proposed an algorithm for computing dynamic slices of concurrent aspect-oriented programs. We have used Concurrent Aspect-Oriented System Dependence Graph (CASDG) as the intermediate representation. Our algorithm is based on marking and unmarking the nodes of the CASDG during run-time. In our algorithm, no new nodes are created in the CASDG during run-time. And the slices generated are correct in nature. Though we have developed our approach for AspectJ language, it can easily be extended for computing dynamic slices of any other AO languages such as AspectWerkz, JMangler, Hyper/J, MixJuice, PROSE, ArchJava and JAC because these tools support AOP with Java.

# Authors

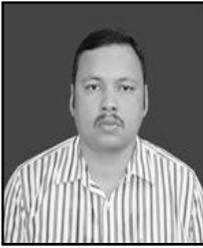

**Abhishek Ray**

Abhishek Ray received his M. E. from Regional Engineering College (now NIT), Rourkela. He served as a faculty of the Department of Computer Science and Engineering at Gandhi Institute of Engineering & Technology, Gunupur, Odisha, from 1998 to 2005. He joined as a faculty in School of Computer Engineering at KIIT University, Bhubaneswar, Odisha in 2005, where he is now Associate Professor. His research interests include software engineering, real-time systems, automata theory and compiler design. He has published five papers in these fields. Mr. Ray has been teaching automata theory and compiler design to UG and PG students at KIIT University, Bhubaneswar for the six years.

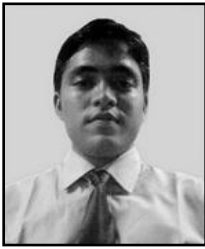

**Siba Mishra**

Siba Mishra received his M.Tech from KIIT University, Bhubaneswar, Odisha, India. His research interests include software engineering, automata theory and discrete mathematics.

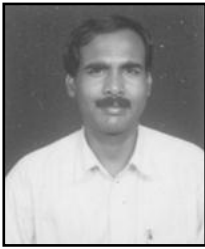

**Durga Prasad Mohapatra**

Durga Prasad Mohapatra received his Ph. D. from Indian Institute of Technology Kharagpur and M. E. from Regional Engineering College (now NIT), Rourkela. He joined the faculty of the Department of Computer Science and Engineering at the National Institute of Technology, Rourkela in 1996, where he is now Associate Professor. His research interests include software engineering, real-time systems, and discrete mathematics and distributed computing. He has published more than thirty papers in these fields. Dr. Mohapatra has been teaching software engineering and discrete mathematics to UG and PG students at NIT Rourkela for the past ten years. He has received many awards including Young Scientist Award for the year 2006 by Orissa Bigyan Academy, Prof. K. Arumugam award for innovative research for the year 2009 and Maharasthra State National Award for outstanding research for the year 2010 by ISTE, New Delhi. He has also received three research projects from DST and UGC, Government of India. Currently, he is a member of IEEE. Dr. Mohapatra has co-authored the book *Elements of Discrete Mathematics: A computer Oriented Approach* published by Tata Mc-Graw Hill.